\documentstyle[aas2pp4]{article}

\def\betiips1{\beta_{i,i+1}}
\def\gamef{{\gamma^{\rm e}}}
\def\gamiips1{{{\gamma^{\rm e}}_{i,i+1}}}
\def\PG{P$\Gamma$}
\def\rhosf{\rho_{\rm SF}}
\def\u{{\bf u}}
\def\VS{V\'azquez-Semadeni}

\newif\ifAMStwofonts
\begin{document}

\title{Is Thermal Instability Significant in Turbulent Galactic Gas?}

\author{Enrique V\'azquez-Semadeni$^1$, Adriana Gazol$^2$ and John Scalo$^3$}

\affil{$^1$Instituto de Astronom\'\i a, UNAM, Apdo. Postal 70-264,
M\'exico D.F., 04510, MEXICO}
\affil{$^2$Instituto de Astronom\'\i a, UNAM, J.\ J.\ Tablada 1006,
Morelia, Mich., 58090, MEXICO} 
\affil{$^3$Astronomy Department, University of Texas, Austin, TX 78712}


%

\begin{abstract}

We investigate numerically the role of thermal instability (TI) as a
generator of density structures in the interstellar medium (ISM), both
by itself and in the context of a globally turbulent medium. We
consider three sets of numerical simulations: {\it a)} flows in the
presence of the instability only; {\it b)} flows in the presence of
the instability and various types of turbulent energy injection
(forcing), and {\it c)} models of the ISM including the magnetic
field, the Coriolis force, self-gravity and stellar energy injection.
Simulations in the first group show that the condenstion process which
forms a dense phase (``clouds'') is highly dynamical,
and that the boundaries of the clouds are accretion shocks, rather
than static density discontinuities. The density histograms (PDFs) of
these runs exhibit either bimodal shapes or a single peak at low
densities plus a slope change at high densities. Final static situations may
be established, but the equilibrium is very fragile: small
density fluctuations in the warm phase require large
variations in that of the cold phase, probably inducing shocks into
the clouds. Combined with the likely disruption of the clouds by
Kelvin-Helmholtz instability (Murray et al.\ 1993), this result
suggests that such configurations are highly unlikely. 

Simulations in the second group show that large-scale turbulent
forcing is incapable of erasing the signature of the TI in the density
PDFs, but small-scale, stellar-like forcing causes the PDFs to transit
from bimodal to a single-slope power law, erasing the signature of the
instability. However, these simulations do not reach stationary
regimes, the TI driving an ever-increasing star formation
rate. Simulations in the third group show no significant
difference between the PDFs of stable and unstable cases, and reach
stationary regimes, suggesting that the combination of the stellar
forcing and the extra effective pressure provided by the magnetic
field and the Coriolis force overwhelm the TI as a density-structure
generator in the ISM, the TI becoming a second-order effect. We
emphasize that a multi-modal {\it temperature} PDF is not necessarily
an indication of a multi-phase medium, which must contain clearly
distinct thermal {\it equilibrium} phases, and that this
``multi-phase'' terminology is often inappropriately used.

\end{abstract}


\section{Introduction}
The idea that thermal instability (TI) plays a fundamental role in
controlling important aspects of the interstellar gas and star
formation in galaxies has been tremendously influential.  As a model
for the formation of cool dense clouds in pressure equilibrium with a
warm rarefied intercloud medium, TI has been invoked to explain the
existence of diffuse interstellar clouds (Field, Goldsmith, \& Habing
1969, FGH), as a means of regulating the mass flow between different
components of the ISM and the star formation rate (Chieze
1987; Parravano 1989), as a major factor allowing star formation to
occur at an appreciable rate in galaxies (Norman
\& Spaans 1997; Spaans \& Norman 1997), the fragmentation of protoglobular
clusters (Murray \& Lin 1996), and cooling flows around luminous cluster
elliptical galaxies and more general situations (see Nulsen \& Fabian
1997; Allen \& Fabian 1998;  Mathews \& Brighenti 2000 for recent discussions).

The linear stability analysis for TI was first worked out in detail by
Field (1965), who clearly delineated the isobaric, isochoric, and
isentropic modes, and examined effects such as magnetic fields and
conduction.  Subsequent work generalized the anlaysis to include
chemical reactions (Yoneyama 1973), and refined the calculation of
relevant heating and cooling rates (see Wolfire et al.\ 1995 for a
recent discussion).  In its simplest form, the isobaric mode of TI
occurs when a 
compression leads to such enhanced cooling that pressure in the
compressed region decreases.  This leads to a picture of dense clouds
in pressure equilibrium with their surroundings, the basis for the
``two-phase" ISM model of FGH.  This conceptual framework was extended
by McKee \& Ostriker (1977) to include a third, hot, phase
representing the interiors of expanding supernova remnants, but still
assumes that clouds form by TI.

The idea of cloud formation by TI suffers from a serious defect: the
linear stability analysis assumes a static uniform initial medium.  It
is now observationally evident that the ISM in galaxies is in a highly
agitated, ``turbulent" state, and so it is questionable whether the
static initial conditoins are ever realized.  The basic question is
whether dynamical motions of the gas (shocks, cloud collisions, or the
turbulence itself) will disrupt incipient clouds faster than they can
condense.

Even within the context of linear stability analyses, there have been
several suggestions that gravitationally induced convective motions
could suppress TI (e.g. Balbus \& Soker 1989).
Numerical studies of the nonlinear evolution of TI in the solar
chromosphere transition region and corona (Dahlberg et al.\ 1987;
Karpen et al. 1988) and in galaxy cluster cooling flows (Loewenstein
 1989; Malagoli et al.\ 1990; Hattori \& Habe 1990; Yoshida et al.\ 1991)
have generally confirmed this suspicion.  The motion of the incipient
condensations leads to vortices, or to complete disruption by
Kelvin-Helmholtz and Rayleigh-Taylor instabilities, although a
magnetic field can diminish the effect in these convective cases.
Moreover, Murray et al. (1993) have showed in some detail how the motion of a
condensation initially in a two-phase medium in pressure equilibrium can
easily lead to disruption of the condensation by dynamical instabilities,
similar to the results quoted above.  Strongly self-gravitating
condensations can avoid this fate, but there is no reason to assume that
such condensations were formed by thermal instability; supersonic turbulence
interactions are sufficient to produce such structures (e.g.,
\cite{PaperIII}). 

What about thermal instability in a supersonic turbulent medium, like
the gas in galaxies?  (By ``turbulent" we mean a disordered but not
completely random velocity [and density] field covering a large range
of scales.)  In the ISM of the Milky Way there is certainly strong
evidence for supersonic motions at all scales above at least 0.1 pc,
in every sort of environment.  Supersonic spectral linewidths are
observed even in regions with no detectable internal star formation
(e.g. the Maddalena molecular cloud complex; see Williams \& Blitz
1998) and in diffuse mostly-H I clouds in which self-gravity is
unimportant (known since the 1950s; see Heithausen 1996 for a recent
study of a subclass of such clouds), as well as in the more
intensively-studied star-forming molecular cloud structures.

A simple argument might suggest that thermal instability may be
inoperative, or at least less efficient, as a cloud formation process
in a supersonically-turbulent ISM.  The isobaric mode of TI in a
region of size $L$ condenses on a characteristic time scale $L/c$,
where $c$ is the sound speed, while the turbulence shears (or expands,
or compresses) the forming condensations on a crossing time scale
$L/v$, where $v$ is the characteristic turbulent speed at scale $L$.
If the turbulence is supersonic, the incipient condensation should be
disrupted faster than it can condense.  However this argument only
applies in an average sense; a region of incipient condensation might
avoid disruptive turbulent interactions for an unusually long
time. More importantly, as we show in sec.\ \ref{pure_force}, when one
considers that the instability growth time decreases with scale size
while the crossing time increases with scale, the above argument can
only apply at sufficiently large scales, and below some size scale the
instability could still operate. Moreover, Hennebelle \& P\'erault
(1999) have recently shown that TI can be {\it triggered} by a
dynamical compression in an originally stable flow.

A quantitative result suggesting that the idea of pressure-confined
clouds, central to models of the ISM which rely on thermal
instability, as well as other quasi-static conceptions, is not viable
in a supersonically turbulent medium was given, using numerical
simulations, by Ballesteros-Paredes, \VS\ \& Scalo (1999, herafter
BVS99). They showed, partly through the evaluation of volume and
surface terms in the virial theorem, that the importance of the
kinetic energy surface terms implies that clouds cannot be considered
as quasi-permanent entities with real ``boundaries," but are instead
continuallly changing, forming and dissolving.  Not only is thermal
pressure incapable of confining turbulent density fluctuations, but
the idea of external turbulent-pressure confinement seems
self-defeating, since the external turbulent stresses mostly serve to
distort and disrupt clouds.

There are additional problems with the two- or three-phase ISM models.
One involves time scales.  The time scale on which a given parcel of the
ISM is subjected to impulsive perturbations, whether by heating or by
shock interactions, may be smaller than the time scale for TI, and
than the time 
for readjustment to pressure equilibrium.  That clouds must be
subjected to frequent stochastic heating events was the basis for
Kahn's (1955) early cloud-cloud collision model for the temperature
distribution of clouds, and the ``time-dependent ISM" models proposed
by Bottcher et al.\ (1970) and examined in detail by Gerola, Kafatos,
\& McCray (1974).  The latter work only included the influence of
stochastic heating and ionization sources (no hydrodynamics), but the
resulting statistical description of the diffuse ISM (Gerola et
al. 1974) was in better agreement with available observations than the
static two-phase models.  A recent summary of (mostly observational)
arguments against the specific three-phase McKee \& Ostriker model is
given by Elmegreen (1997).  Moreover, the conclusion that pressure
equilibrium is unlikely because of the frequency of either cloud
collisions or shocks from supernovae or cluster superbubbles has been
found independently several times in the literature (e.g. Stone 1970;
Heathcote \& Brand 1984; Bowyer et al.\ 1995; Bergh\"ofer et al.\ 1998;
Kornreich \& Scalo 2000).

Given all the above results, it is important to examine whether the
traditional picture of a multi-phase ISM structured by TI should be
retained as a useful model.  By ``multi-phase'' we are referring to a
medium containing various different thermodynamic {\it equilibrium}
regimes, some of them possibly stable and others unstable, with the
requirement that a multi-phase medium involves fluid parcels
undergoing a ``phase transition'' when transiting between these
equilibria. The structuring of the density field by the multiphase
nature of the medium should be reflected in multimodal density and
temperature pdfs, with gas accumulating in the stable ``phases''.

However, note that the temperature field is expected to show a bimodal
PDF even in the absence of TI, simply because of the functional shape
of the interstellar cooling curve, which has the form of two plateaus
separated by a sharp discontinuity at $\sim10^4$ K (see Dalgarno \&
McCray 1972, Fig.\ 11). For example, the simulations of Korpi et al.\
(1999) do not contain real phases, since no background heating capable
of balancing the cooling is included; yet, the temperature still
exhibits a bimodal distribution in those simulations.  But for a
TI-structured multi-phase ISM, the gas must be concentrated into two
(or more) density components: dense clouds and a rarified intercloud
medium. Two-dimensional simulations of the ISM including a typical
cooling function, but containing no TI (Scalo et al.\ 1998) showed
no indication of a bimodal density PDF. For this reason in this paper
we consider a bimodal density distribution as the signature of a
TI-structured ISM. Actually, our experiments have shown that, under
certain conditions, the PDFs of unstable flows may be unimodal, but with
an added slope change where the second peak should be located. Thus,
we consider either of these PDF shapes as a signature of density
structuring by TI.

It is also of great interest to investigate whether TI can
significantly enhance the formation of clouds by turbulent
 interactions.  As shown in V\'azquez-Semadeni, Ballesteros-Paredes 
 \& Rodrigu\'ez (1997), BVS99  and
 references therein, ``clouds" naturally form in a supersonic
turbulent medium, without any need for instabilities, as turbulent
velocity streams interact to form compressed layers, as suggested by
Elmegreen (1993).  

The present work attempts to investigate these ideas using numerical
simulations in two dimensions.  First we simulate the evolution of an
initially static gas whose cooling and heating functions should
guarantee TI, and show that, as expected, the density distribution 
shows the signature of the instability. Second, we perform
simulations of a randomly forced turbulent flow in the presence of the
instability, showing that large-scale forcing does not seem to be able 
to prevent the development of the instability, but small-scale forcing 
does. 

Finally, a third series of simulations considers a supersonic
turbulent gas, including self-gravity, the magnetic field, the
Coriolis force, and energy input due to star formation, in stable,
marginal and unstable cases.  We demonstrate that in the latter case
the density distribution does not show the signature of the
instability. The temperature and density are still anticorrelated, as
expected from the cooling function, but a two-phase density field is
not realized.  This is ultimately due to the fact that, when the
cooling time scale is much smaller than the hydrodynamic time scale, the
density is ``slaved" to the dynamics, not the temperature (cf., \VS,
Passot \& Pouquet 1995); another way of expressing this is that the
density field is controlled by ram pressure/advection, not the thermal
pressure.

\section{The Model}\label{sec-mod}
We use the numerical ISM model presented by Passot, V\'azquez-Semadeni
\& Pouquet (1995), which uses a single-fluid approach to describe the
ISM on a 1-kpc$^2$ plane on the Galactic disk, centered at the solar
Galacto-centric distance. In this model, various physical ingredients
can be included at will, such as self-gravity, the magnetic field,
disk rotation, as well as model terms for the radiative cooling
($\Lambda$), a diffuse background radiation ($\Gamma_{\rm d}$), and
energy input due to star formation ($\Gamma_{\rm s}$). In this paper,
we solve the equations in logarithmic form for the density and
temperature, as done in Gazol-Pati\~no \& Passot (1999). This is
because the shocks induced by the development of the instability are
quite strong, and the logarithmic approach avoids the appearance of
negative values of the density and temperature due to the Gibbs
phenomenon in the vicinity of discontinuities, giving somewhat more
robustness to the code. The equations solved read
\begin{eqnarray}
&&
\quad\frac{\partial\ln \rho}{\partial t}+\nabla\cdot(\ln\rho\bf u)
      +(1-\ln\rho)\nabla\cdot {\bf u}
\nonumber\\ 
&&
=\mu(\nabla^2\ln\rho +
      (\nabla\ln\rho)^2) ,\label{mhd1}\\
&&
\quad\frac{\partial{\bf u}}{\partial t}+{\bf u}\cdot\nabla{\bf u}=
\nonumber\\
&&
     -\frac{\nabla p}{\rho} -{(\frac{J}{M})}^2\nabla\varphi
     -\nu _8\nabla ^8{\bf u} +\nu_2 (\nabla^2 \u + \frac{1}{3} \nabla
     \nabla \cdot \u)
\nonumber \\ 
&&
+\frac{1}{\rho}(\nabla\times{\bf B})\times{\bf B}
 -2\Omega\times{\bf u} \label{mhd2}\\
&& 
\quad\frac{\partial\ln e}{\partial t}
    +{\bf u}\cdot\nabla\ln e =
     -(\gamma -1)\nabla\cdot{\bf u}
\nonumber\\
&&
+\frac{\kappa _T}{\rho}
      (\nabla ^2\ln e +(\nabla\ln e )^2)
+\frac{\Gamma_{\rm d}+\Gamma_{\rm s}+
      \rho \Lambda }{e},
      \label{mhd3}\\
&&
\quad\frac{\partial{\bf B}}{\partial t}=
     \nabla\times ({\bf u\times B})-\nu _8\nabla ^8{\bf B}
  + \eta \nabla^2 {\bf B}\label{mhd4}\\
&&
\quad\nabla ^2\varphi =\rho -1,\label{mhd5}\\
&&
\quad P=(\gamma - 1)\rho e,\label{mhd6}
\end{eqnarray}
where $\rho$ is the fluid density, ${\bf u}$ the velocity, $P$ the
thermal pressure, $\varphi$ the gravitational potential, $\Omega$ the
angular velocity due to Galactic rotation, ${\bf B}$ the magnetic
field and $e$ the specific internal energy, related to the temperature
$T$ by $e=c_vT$, $c_v$ being the specific heat at constant
pressure. In eq.\ (\ref{mhd2}), $J\equiv L_{\rm o}/L_{\rm J}$ is the
Jeans number, measuring the length of the integration box in units of
the Jeans length, and $M \equiv u_{\rm o}/c_{\rm o}$ is the Mach
number of the velocity unit $u_{\rm o}$ with respect to the
temperature unit, given by its corresponding speed of sound $c_{\rm
o}$.  In eq.\ (\ref{mhd3}), $\gamma$ is the ratio of specific heats at
constant pressure and volume, and $\kappa _T$ is the thermal
diffusivity. 

Momentum and magnetic field dissipation are included by a
combination of a quadrilaplacian ``hyperviscosity'' operator which
confines the viscous effects to the smallest scales, and a second
order operator, which filters out possible oscillations in the
vicinity of strong shocks. The use of the hyperviscosity operator
allows the use of much smaller (typically by a factor of 10) values of the
second-order diffusivities than otherwise. An artificial mass
diffusion term in the continuity equation has been added in order to
smooth out the density gradients, allowing the code to handle stronger
shocks. Its effect is only to spread out and somewhat compress density
peaks. The effect of this term has been quantified by \VS,
Ballesteros-Paredes \& Rodr\'iguez (1997) and BVS99.

The physical units are $L_{\rm o}=1$ kpc,
$\rho_{\rm o}=1 m_{\rm H}$cm$^{-3}$, $u_{\rm o}=11.7$km s$^{-1}$,
$t_{\rm o}=1.3 \times 10^7$ yr, $T_{\rm o}=10^4$ K, and $B_{\rm o}=5
\mu$G. Throughout the paper, we omit $m_{\rm H}$, the proton mass, in
the specification of densities.

The above equations are solved in two dimensions at a resolution of
$128^2$ grid points (implying a spatial resolution of 7.8 pc) by means
of a pseudo-spectral method, which imposes
periodic boundary conditions. The temporal scheme is a third order
Runge-Kutta for the non-linear terms, combined with a Crank-Nicholson
for the linear terms. The initial conditions are set up in Fourier
space, uncorrelated in all variables, and characterized by a power
spectrum of the form $k^4\exp\left[-2\left(k/k_0\right)^2\right]$,
with random phases. In the case of the magnetic field, the initial
condition involves a uniform component of strength $B_1=1.5
\mu$G in the $x$-direction mimicking the
azimuthal component of the field on the Galactic plane, upon which a
fluctuating field of rms strength $5 \mu$G is added.

An important note is that, in spite of the usage of logarithmic
variables and hyperviscosity, our simulations cannot handle very strong
shocks because of the spectral scheme used (the Gibbs phenomenon
produces artificial oscillations near discontinuities), and
therefore the simulations must be stopped when the gradients of the physical
variables become too steep. However, they are able to reach advanced
enough stages of the development of the instability that its fully
dynamical character is seen, and the structures arising are well
defined. 

We describe below only those model terms which are directly related with
simulations presented in the following sections. Further details
concerning the  choice of the other terms and parameters can be found in 
Passot et al.\ (1995) and \VS\ et al.\ (1996). Table 1 presents the relevant
parameters for the runs in this paper.
\subsection{Radiative cooling and diffuse heating}
The heating ($\Gamma$) and cooling ($\rho \Lambda$) rates are {\it per
unit mass}, and thus differ somewhat from the standard definitions,
which are per unit volume. In eq.\ (\ref{mhd3}), the term $\Gamma_d$
models the heating of the gas due to photoelectrons ejected from dust
grains due to the background UV radiation field, or heating by
low-energy cosmic rays, and is taken as a constant $\Gamma_d=\Gamma_0$
everywhere and throughout the duration of the simulations.

The simulations use a cooling function 
with piecewise power-law dependence on the temperature of the form
\begin{equation}
\Lambda =C_{i,i+1}T^{\beta_{i,i+1}}\:\:\:{\rm for}\:\:\:T_i\leq T<T_{i+1},
\label{cool}
\end{equation}
The various runs have different values of $\beta_{i,i+1}$, given
in Table 1. One set of values of these exponents, namely $\beta_{12}=1$,
$\beta_{23}=0.25$, $\beta_{34}=4.18$ and $\beta_{45}=-0.53$, was chosen as a
reasonable piecewise power-law fit to the radiative cooling curve of
\cite{DaMc72} for a fractional ionization of $10^{-4}$ at $T<10^4$ K
(see also \cite{spitzer}). For this fit, we have chosen the
transition temperatures $T_i$ as $T_1=10$ K, 
$T_2=398$ K, $T_3=10^4$ K, $T_4=10^5$ K and $T_5=10^7$ K. As discussed 
below, this particular combination of values of $\beta_{i,i+1}$ and $T_i$
give a thermally unstable range (isobaric mode) between $T_2$ and
$T_3$. Other runs 
have variations of this cooling function in order to obtain thermal
stability or marginal instability in the range $398-10^4$ K
without a significant increase of the cooling rate above $10^4$ K. Due to
continuity reasons, the values of coefficients $C_{i,i+1}$ must also be
changed, and are also listed in Table 1. Finally, all runs share the
same values of $\beta_{45}$ and of $T_1$, $T_3$, $T_4$ and $T_5$, and are
therefore not listed in Table 1.

As discussed in previous papers (e.g., \cite{PaperI}; \cite{PaperIII}),
the thermal time scales are typically much shorter than the dynamical
ones, implying that the turbulent motions are quasi-static compared to 
the background heating and cooling rates, and allowing for the
establishment of thermal equilibrium. With the adopted power-law
behavior for these processes, the equilibrium temperature and pressure are
\begin{eqnarray}
&&
T_{eq}=\left(\frac{\Gamma_0}{C_{i,i+1}\rho}\right)^{1/\beta_{i,i+1}}
\label{thequ1}\\ 
&&
P_{eq}=\frac{\rho^{\gamef_{i,i+1}}}{\gamma}
\left(\frac{\Gamma_0}{C_{i,i+1}}\right)^{1/\beta_{i,i+1}}, \label{thequ2}
\end{eqnarray}
where the (piecewise) {\it effective polytropic index} is given by
$\gamef_{i,i+1} \equiv 1-1/\beta_{i,i+1}$.  The isobaric mode of the
TI, characterized by a decrease in the pressure as the density
increases, develops when $\gamiips1<0$, corresponding in this case to
$\betiips1<1$. The lower pressure causes an effective ``suctioning''
action by the high-density regions. The values of the resulting
$\gamef_{12}$ and $\gamef_{23}$ are also given in Table 1. The value
of $\gamef_{34}$ is always taken as $\gamef_{34}=0.76$, and therfore
not listed in Table 1. Note that the flow behaves as $P \propto
\rho^\gamiips1$ in the density interval $\rho_i < \rho \leq
\rho_{i+1}$, where the ``transition'' densities $\rho_i$ are defined
as the values whose corresponding equilibrium temperatures (eq.\
[\ref{thequ1}]) coincide with $T_i$.\footnote{Note that runs with
different cooling 
functions must also have different transition densities $\rho_2$ and
$\rho_3$. It is not possible to vary the cooling and heating functions
in such a way as to keep both the transition densities and temperatures
constant simultaneously, without varying the effective rates
excessively.}  Note also that some of the runs,
in addition to having a thermally unstable range with $\gamef_{23}<0$,
have a marginally stable regime at low temperatures, with
$\gamef_{12}=0$. This is the case in particular for our fit of the
\cite{DaMc72} cooling curves, and its consequences are discussed in \S
\ref{pure_morph}.

The dispersion relations for the combined thermal+gravitational
instability in the presence of a uniform field $B_1$ have been derived
for the two-dimensional case by Passot et al.\ (1995) (for the general 
3D case, see Elmegreen 1994), and read:
\begin{eqnarray}
\omega_{\rm t}^2 = J^2 - k^2 \bigl({\gamiips1\over\gamma}T_{\rm eq} +
B_1^2\bigr) - \kappa^2 \label{disp_rel_t} \\
\omega_{\rm l}^2 = J^2 - k^2 {\gamiips1\over\gamma} T_{\rm eq} -
{\kappa^2 \omega_{\rm l}^2\over \omega_{\rm l}^2 + k^2 B_1^2},
\label{disp_rel_l} 
\end{eqnarray}
where $\omega_{\rm l}$ and $\omega_{\rm t}$ respectively denote the
growth rates for perturbations in the longitudinal and transverse
 directions with respect to $B_1$, $k$ denotes the wavenumber, $J$ is
the Jeans number 
defined above, and $\kappa=2\Omega[1 + (1/2)(y/\Omega)(dy/d\Omega)$ is the
 epicyclic frequency. Unstable wavenumbers are those for which
$\omega^2>0$. For 
the values of the parameters used in the unstable runs, typical unstable
ranges are $k \gtrsim 1$, implying that essentially all scales
included in the simulations are unstable. Note that the thermal
instability ($\gamef < 0$) has the effect of {\it reversing} the
instability criterion with respect to the purely gravitational case,
in the sense that now it is scales {\it smaller} than some threshold
value that are unstable.

Table 1 also contains information about the thermal equilibrium regime
for the heating and cooling functions used here, in particular
$\rho_2$ and $\rho_3$, and the equilibrium temperature
$T(\langle\rho\rangle)$ at the mean density, $\langle \rho
\rangle=1\;{\rm cm}^{-3}$. Note that the value of $\Gamma_0$
used here is in general not the same as that used by \VS\ et al.\ (1995).
We have chosen $\Gamma_0$ in order to ensure that
$T(\langle\rho\rangle)$ lies within the unstable range.
\subsection{Forcing} \label{forcing}
We consider two kinds of forcing (i.e., energy injection) for the
turbulent runs. The first is a large-scale forcing, accomplished through an
additional acceleration term in the momentum equation (eq.\
[\ref{mhd2}]), taken as a random vector field with a 
spectrum proportional to $k^4$ below the forcing wavenumber $k_{\rm for}=4$, 
and $k^{-4}$ above $k_{\rm for}$. We consider two subcases of this forcing,
one with 100\% solenoidal (rotational) components and the other with 50\%
solenoidal and 50\% compressible components.

The second type of forcing is applied at small scales, and consists of
local, discrete heating sources turned on at a grid point $x$ whenever
$\rho(x)>\rhosf$ and $\nabla\cdot u(x)<0$, where $\rhosf$ is a free
 parameter, but taken equal to $6
\langle\rho\rangle$ always. 
The sources stay on for a time interval $\Delta t=6\times
10^6$yr. This forcing mimics the effect of ionization heating from
massive stars, and acts on the scale of $\sim 5$ pixels
(\cite{PaperI}; \cite{PaperII}).

\subsection{The simulations}\label{simulations}

In order to investigate the interplay between the TI and the
turbulent dynamics, we have performed three sets of three simulations
each. First, we have considered the development of the instability
alone (runs labeled ``pure'' or, simply, ``P'' in Table 1), by
simulating a purely hydrodynamic flow without the magnetic field,
rotaion or self-gravity. Only the appropriate cooling and background
heating functions necessary for the existence of a thermally unstable
range are included. The three runs differ in details of the cooling
and background heating functions, chosen to show the variety of
structures that develop and the effects on them
of the closeness of the transition densities to the mean density. The
label ``P'' is thus followed by a mnemonic indicating the value of
$\Gamma_0$ used in each run.

The second set of simulations (referred to as the ``hydrodynamic'' runs, or
``H'' in Table 1) are similar to the P runs, except that they include a 
variety of forcing schemes, as described in \S \ref{forcing}. These
are designed to test the effect of the scale and the amount of
compressible content of the energy injection on
the supression of the instability by the turbulence. As indicated in
\S \ref{forcing}, the small-scale forcing is stellar-like, i.e., in
the form of localized sources of heating which create expanding
bubbles of hot gas, while the large-scale forcing is random, with
varying ratios of compressible-to-solenoidal content.

Finally, we have performed three ISM-like simulations (labeled ``ISM''
in Table 1), including the magnetic field, rotation and self-gravity,
as well as the stellar-like forcing. The runs differ from each other
in whether the cooling/heating combinations produce a stable, marginal
or unstable regime, and are designed to test the effects of the
presence of the TI in an ISM-like medium.

For the ``pure'' and ``hydro'' runs we have used initial fluctuations
in the density and temperature of rms amplitude equal to 0.1 times
their mean values ($=1$), with zero initial velocity fluctuations. For
the full ISM-like runs, we have used initial fluctuations with rms
amplitudes equal to the mean values of the variables, except for the
magnetic field, whose fluctuations are taken as described in \S
\ref{sec-mod}.

\section{Results and discussion}\label{sec-res}

\subsection{Development of the pure instability}\label{pure}

\subsubsection{Morphology and dynamics} \label{pure_morph}

\begin{figure}
\plotone{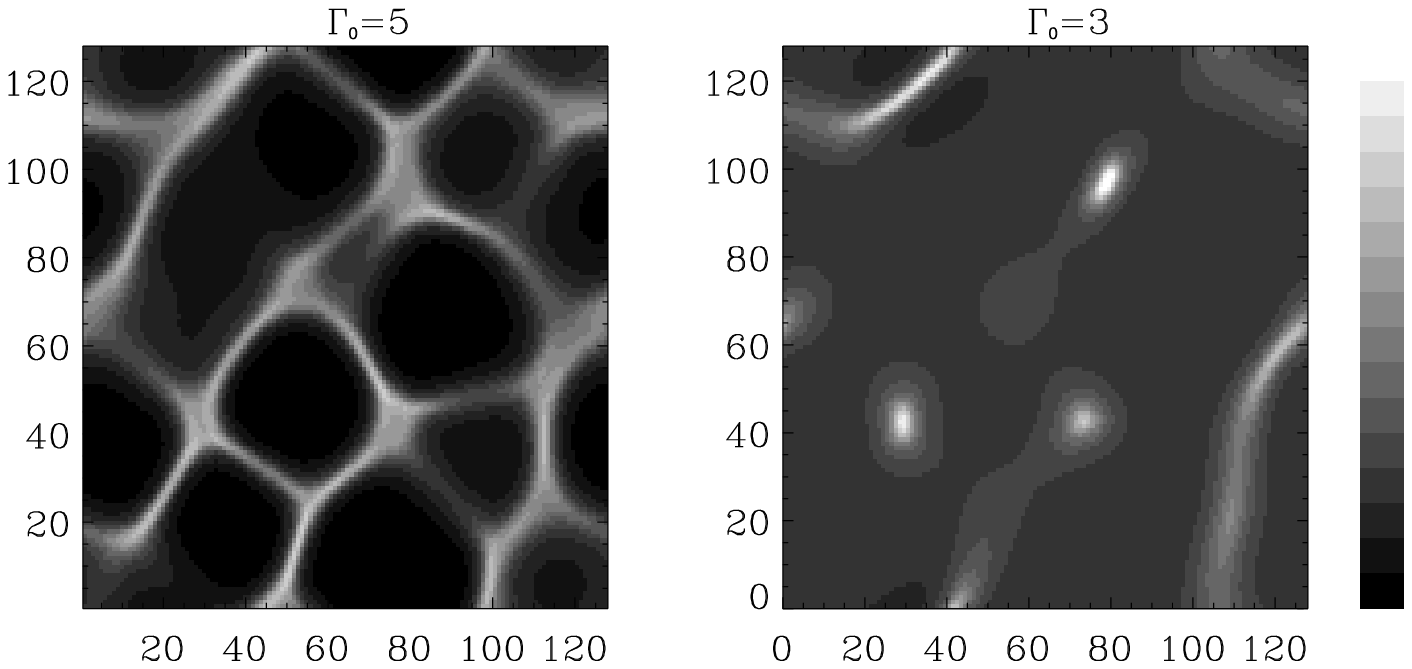}
\caption{Density field of runs a) \PG 5 ({\it left}) and b) \PG 3
({\it right}). In these runs, only the TI is at work, starting from
small density fluctuations. Run \PG 5 is seen to develop a network of
filaments, while run \PG 3 transits directly to the formation of
isolated density structures. }
\label{G5G3}
\end{figure}

The density fields for runs \PG 5 and \PG 3 at an advanced stage of
the instability development are shown in Figs. \ref{G5G3}a (left
panel) and \ref{G5G3}b (right panel), respectively. In run \PG 5
(\ref{G5G3}a), the gas is seen to evolve into a ``bee-nest'' structure
formed by a network of dense ($\sim 10\;{\rm cm}^{-3}$), cold ($\sim
50\;{\rm K}$) filaments enclosing warm ($\sim 10^4\;{\rm K}$) and
rarefied ($\sim 0.3\;{\rm cm}^{-3}$) cells. The cells' shape and size
are determined by the initial conditions. In this run we have taken
$k_0=2$, implying that the characteristic scale of the initial
fluctuations is one half of the box size. Note that the TI leads to
the collapse of regions containing a density excess, whose
characteristic size is one fourth of the integration domain. The
roundish blobs formed at intersections between filaments originate
from density maxima, while the dense filaments can be traced back to
``saddle points'' located between two density maxima and two
minima in the initial conditions.\footnote{The initial conditions for
the 2D density field are mosaics of small-amplitude, alternating
maxima and minima. Typically, imaginary lines joining two
maxima and lines joining two minima intersect, forming an ``$\times$''
pattern, at the center of which there is a saddle point.} The
evolution of this simulation is 
reminiscent of ``pancake'' formation due to the gravitational
instability in cosmological simulations of large-scale structure
formation (e.g., Padmanabhan 1993). However, this bee-nest structure
is transient, as we discuss below.

\begin{figure}
\plotone{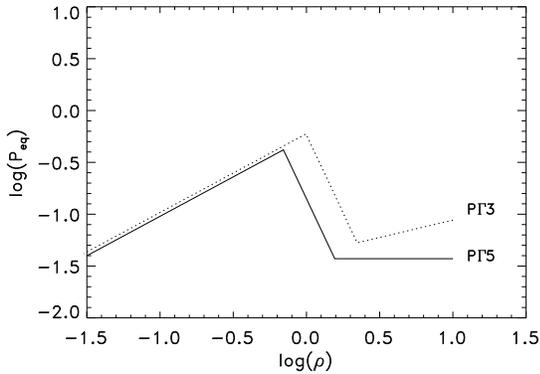}
\caption{Equilibrium pressure as a function of density as given by
eq.\ (\ref{thequ2}), for the cooling curves used in runs \PG 5 ({\it
solid line}) and \PG 3 ({\it dotted line}).}
\label{pvsrho_th}
\end{figure}

\begin{figure}
\plotone{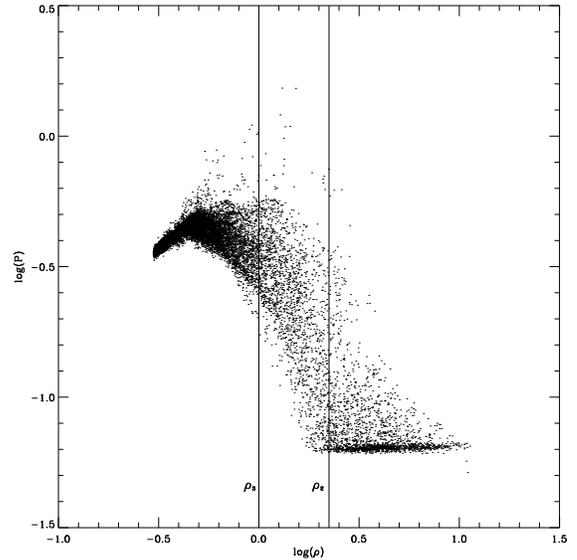}
\caption{A pixel-by-pixel plot of pressure {\it vs.} density for run
\PG 5. Compare to the solid curve in fig.\ \ref{pvsrho_th}. The
thickness of the negative-slope portion is due to the finite cooling
times, which cause the temperature of a fluid parcel to depend not
only on its instantaneous density, but also on the speed at which it
is varying. Numerical noise due to the abrupt transition from one
cooling regime to another also contributes to the scatter.}
\label{G5Pvsrho}
\end{figure}

The condensation process in principle stops after the pressures in the 
warm and the dense (cold) phases equalize and any remaining kinetic energy
due to the inertia of the gas is dissipated. This implies that the
mass redistribution induced by the instability does not end when the
density has ``crossed'' the transition values, but at more disparate
values of the density such that the pressures in the two phases are
equal. For example, fig.\ \ref{pvsrho_th} shows the equilibrium
pressure, as given by eq.\ (\ref{thequ2}) for runs \PG 5 and \PG 3. It
can be seen that in order for the pressure in the warm phase to equal
that of the dense phase, a density of roughly 0.03 cm$^{-3}$ is
required in the former. (Note that, since run \PG 5 has $\gamma_{12}=0$,
the equilibrium can only be reached by decreasing the warm phase'
density.) However, at the final time reached by run \PG 
5, the pressure in the warm phase is still much larger than in the
dense phase, as can be seen in fig.\ \ref{G5Pvsrho}. That is, this run
stopped only because the shocks became too strong for the code to
handle, but it was still in a highly dynamical contracting stage.

\begin{figure}
\plotone{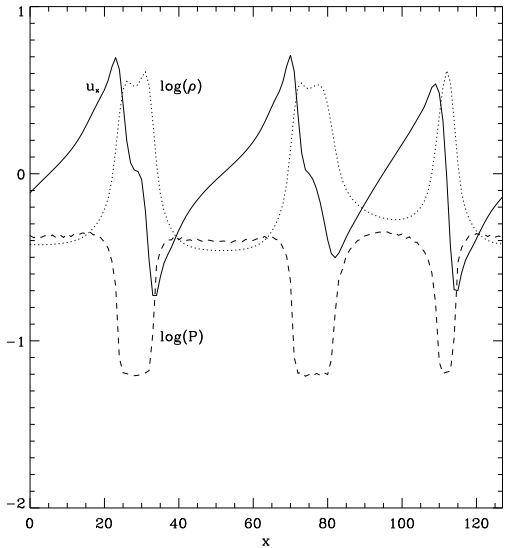}
\caption{Horizontal cuts through run \PG 5 at $y=42$ of the density
({\it dotted line}), $x$-velocity ({\it solid line}) and thermal
pressure ({\it dashed line}). The two structures on the left are
blobs (see fig.\ \ref{G5G3}a), and the one on the right is a
filament. Double-shock structure characteristic of
shock-bounded slabs, with a central plateau bounded by abrupt jumps,
is seen in all three variables in the blobs. It cannot be seen in
the filament on the right because this structure is too small, and
viscosity blends the two shocks in a single one.} 
\label{cuts}
\end{figure}

The highly dynamical nature of the condensation process
is clearly seen in fig.\ \ref{cuts}, which shows
horizontal cuts at $y=42$ through the (log of the) density (dotted
line), (log of the) pressure (dashed line), and the $x$-velocity
(solid line) fields of run \PG 5 (fig.\ \ref{G5G3}a). The cuts go
through two blobs and one filament. A double-shock structure
is observed in the blobs, which are the two-dimensional equivalent of
one-dimensional shock-bounded slabs. This should also be the case of the
filament, but it is too thin for the code to resolve the double-shock
structure, since shocks in the code are spread over roughly five grid
points by viscosity. 

\begin{figure}
\plotone{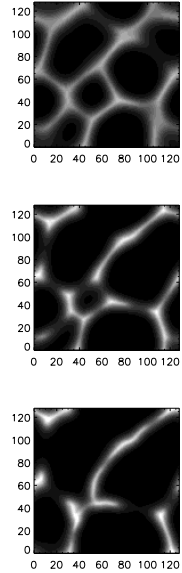}
\caption{Three snapshots of the density field of run \PG .25, at
$t=12.4$ ({\it top}), $t=16.4$ ({\it middle}) and $t=19.3$ ({\it
bottom}). The times are in code units, with 1 code time unit = 13
Myr. Note that the filament network fragments and starts producing
isolated entities.}
\label{G.25t_dens}
\end{figure}

In order to see the subsequent evolution, we thus performed another
run, named \PG .25 with slower heating and  
cooling rates (by decreasing the coefficients) but with the same
temperature dependence as run \PG 3 (except for a slight change in
$T_2$), which therefore evolved more slowly, developing milder shocks, and
allowing us to reach more advanced stages of the instability
development. A temporal sequence in the evolution of this run is
shown in fig.\ \ref{G.25t_dens}. It is seen that this run also develops 
the bee-nest structure, but later the filaments disrupt and are accreted into
the actual peaks. Moreover, some of the cells ``implode'', and their
surrounding filaments and peaks collapse into the void, due to the
small pressure differences between cells, traceable to the randomness
of the initial conditions. The behavior thus continues
to resemble that induced by gravitational instability in cosmological
simulations, as the pancakes later collapse into isolated structures
(e.g., Padmanabhan 1990). 

Run \PG 3 (fig.\ \ref{G5G3}b), on the other hand, does not show the
bee-nest structure of run \PG 5, but shows roundish
blobs within elongated, isolated structures, and only one truly
filamentary structure. More importantly, run \PG 3 never develops 
the thin, dense filaments seen in run \PG 5. Instead, only thick,
moderate-density bands appear between the peaks, and then
disappear. We attribute this difference in part to the fact
that in run \PG 3, $\rho_3$ is very close to the mean
density (see Table 1). This causes any initial underdensity to quickly
reach the stable range, implying that a relatively large
amount of mass remains in the warm phase (note that the buildup of an
underdensity entails the evacuation of that region). 

\begin{figure}
\plotone{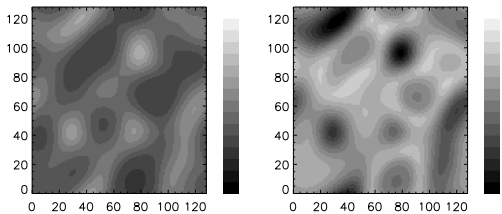}
\caption{Images of the density ({\it left}) and thermal pressure ({\it
right}) for run \PG 3 at an intermediate stage of the instability
development. White means large values and black means low values.
Although the voids have larger pressures than the density
peaks, the largest pressures occur at the interfaces between voids and 
filaments or peaks. Filaments have larger pressures than the voids,
and therefore ``dissolve'' rapidly. The presence of such high-pressure
interfaces is expected in general in the development of the
instability if the density field is not perfectly smooth.}
\label{G3rho_P}
\end{figure}

However, the main reason for the non-appearance of the filaments seems 
to be again the fact that the densities of the warm and cold phases
required for equilibrium are significantly more distant from the mean
density than $\rho_2$ and $\rho_3$. This implies that,
during the development of the instability,  
the pressure maxima do not occur at the centers of the voids, but in
the boundaries between the voids and the high density regions (fig.\
\ref{G3rho_P}). In the particular case of run \PG 3, $\rho_3$ is so
close to $\langle \rho \rangle$ that the voids reach
smaller pressures earlier than the unstable filaments. The latter then
reverse their condensation process, loosing their mass to both 
the voids and the peaks, and ultimately merging with the intercloud
medium. The peaks do not suffer this ``rebound'' because, due to their
larger initial densities, they reach lower pressures faster than the
voids. 

Even though the above effect is particularly noticeable in run \PG 3, we
expect it to apply in general. Therefore, unless the density field is
perfectly uniform within the phases, and the interface between them is
a perfect discontinuity, a larger pressure is expected to be always
present near the interface between the two phases than in either one
of them. {\it This suggests that a steady state with constant pressure
everywhere is difficult to attain.}

\subsubsection{On the final state of the instability} \label{fin_state}

Another consequence of the above discussion is that, since the gas is
a fluid, the boundary of the ``clouds'' that make up the dense phase
are accretion shocks, as discussed by \cite{BVS99}, rather than the
quiescent density jump at constant pressure usually assumed in the
literature. Of course, quiescent pressure equilibrium
may be established after the pressures have equalized, and the
boundary shocks subside, but this may require very long times, and
besides, the balance is very delicate. In general, there is a continuum of
possible equilibria between the dense and rarified phases, obtained by 
equating their pressures, as given by eq.\ (\ref{thequ2}), and
solving for $\rho_{\rm d}$, the density in the dense
phase, as a function of $\rho_{\rm r}$, the density in the rarified
(warm) phase. We obtain
\begin{equation}
\rho_{\rm d}=\left[\left(\frac{C_{12}}{\Gamma_0}\right)^{1/\beta_{12}}
\left(\frac{\Gamma_0}{C_{34}}\right)^{1/\beta_{34}} \rho_{\rm r}^{\gamma_{34}}
\right]^{1/\gamma_{12}}. \label{rho1_rho3}
\end{equation}
The dependence of $\rho_{\rm d}$ on $\rho_{\rm r}$ is thus determined
essentially by the ratio $\gamma_{34}/\gamma_{12}$.  As an example,
plugging in the corresponding parameters for run \PG3, which has
$\gamma_{12}=1/3$, we find that $\rho_{\rm d}=3.2 \times 10^3
\rho_{\rm r}^{2.28}$. Thus, in this case, the equilibrium density of
the dense phase depends sensitively on the density of the warm phase,
minute fluctuations in $\rho_{\rm r}$ requiring large changes in
$\rho_{\rm d}$. For example, at $\rho_{\rm r}=0.1$, $\rho_{\rm
d}=16.8$, while at $\rho_{\rm r}=0.3$, $\rho_{\rm d}= 205$ (all in
units of the mean density). Such large density variations for the
dense phase's density will require highly dynamical adjustments which
will most likely involve new shocks, especially since the sound speed
in the warm phase is much larger than in the dense medium. This,
together with the result by Murray et al.\ (1993) that the condensed
clouds are likely to be set in motion relative to the diffuse phase by 
buoyancy or other effects, and then easily disrupted by
Kelvin-Helmholtz-type instabilities, strongly suggests that the static 
configuration is highly improbable.

In cases where $\gamma_{12}=0$, eq.\ (\ref{rho1_rho3}) does not apply, and
the pressure of the dense phase becomes 
independent of its density. Thus, pressure equalization can only be
accomplished through evacuation of the warm phase until its pressure
drops to the value within the dense gas. In the meantime, the density
of the dense gas may increase by very large amounts, most likely
involving violent compressions. We conclude 
that the static configuration is very hard to realize, in agreement
with the conclusions of \cite{BVS99}.

In any case, the formation of static configurations is possible in
principle, although we have seen here that this is a direct
consequence of the change in the effective polytropic exponent
$\gamef$ from one phase to the other, as discussed in detail by
\cite{BVS99}. This suggests that cores within molecular clouds, which
have essentially the same value of $\gamef$ as their parent clouds,
should not be expected to reach hydrostatic equilibrium.

\subsubsection{PDFS} \label{pure_pdfs}

The density PDF has been the subject of much recent work. \VS\ (1994)
and Padoan, Nordlund \& Jones (1997) have reported lognormal PDFs for
two- and three-dimensional {\it isothermal} flows, respectively. Passot \& \VS\
(1998) discovered that the functional form of the PDF for polytropic
($P \propto \rho^\gamef$) one-dimensional flows depends on the effective
polytropic exponent $\gamef$,
developing a power-law tail at high densities for $0< \gamef <1$ and
viceversa, with a lognormal indeed developing when $\gamef=1$ (see
also Nordlund \& Padoan 1999). Scalo
et al.\ (1998) found that the simulations of the ISM by Passot et al.\ 
(1995), as well as 2D Burgers-like (without thermal pressure) runs
developed power-law PDFs analogous to the $0 < \gamef <1$ 
cases of Passot \& \VS\ (1998). 

In the thermally unstable case, the instability causes the evacuation of 
material from the unstable to the stable regions, producing a
two-phase density field characterized in principle by a
bimodal PDF. In the remainder of this
paper we use the density PDF as a diagnostic of the relative
importance of the instability in the dynamics of the various
simulations.

\begin{figure}
\plotone{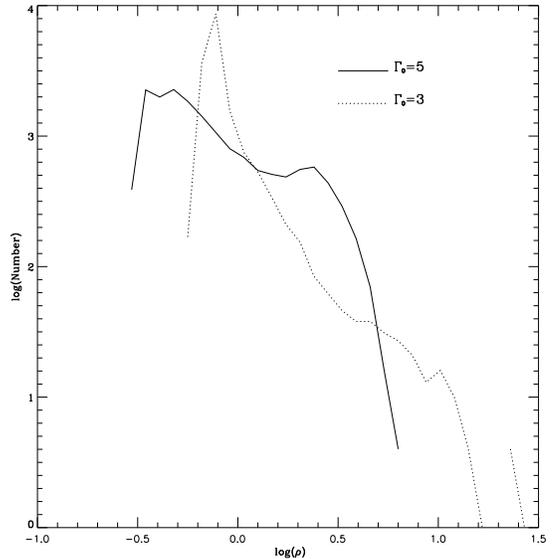}
\caption{Probability density functions (PDFs) or histograms of the
density field for runs \PG 5 ({\it solid line}) and \PG 3 ({\it dotted
line}). The transition densities for both runs are shown as vertical
lines, with the same line type as the histogram for each run. The PDF
of run \PG 5 shows two peaks outside its corresponding transition
densities, while that of run \PG 3 shows only the low-density
peak. Instead of the high-density peak, a slope change beyond its
upper transition density is seen. We consider either kind of PDF a
signatured of the dynamical effect of the TI.}
\label{G5G3pdfs}
\end{figure}

The density PDFs for runs \PG 5 and \PG 3 corresponding to the fields
shown in fig.\ \ref{G5G3} are presented in fig.\
\ref{G5G3pdfs}. The vertical lines denote the transition densities
$\rho_2$ and $\rho_3$ for the two runs. The morphological and
dynamical features described in the last section manifest themselves
in the PDFs as well. In both cases, a peak in the PDF is observed just
below the lower transition density, $\rho_3$. However, in the case of
run \PG 5, $\rho_3$ is not too similar to the mean density, and this
implies that a substantial amount of mass has to be transferred to the 
dense phase. As a consequence, a noticeable peak appears in the PDF
largewards of the upper transition density, $\rho_2$. Instead, for run 
\PG 3, much of the mass remains in the low density phase, and there is 
no clear peak above $\rho_2$. Only a change in the slope of the PDF is 
seen there. 

\begin{figure}
\plotone{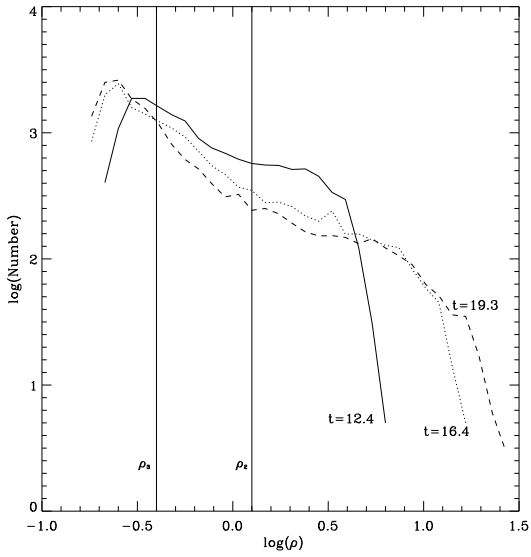}
\caption{Evolution of the density PDF of run \PG .25. The PDFs are
shown at the same times as the fields shown in fig.\
\ref{G.25t_dens}. The PDF is initially bimodal, but, as the
instability continues to develop, the high-density peak is stretched
out towards even higher densities, and becomes simply a flatter
portion of the PDF.}
\label{G.25t_pdfs}
\end{figure}

Furthermore, the PDF for run \PG 3 seems to extend to much higher
densities than that of run \PG 5. We interpret this as a consequence
that run \PG 3 reached a more advanced stage in the development of the 
instability than run \PG 5, so that densities much higher than the
transition value 
are reached as the densities in each phase approach the equilibrium
values. In fact, this phenomenon should have also been observed in
run \PG 5, had it reached more evolved stages. Indeed, in fig.\
\ref{G.25t_pdfs} we show the density PDF at three different times for run
\PG .25 which, as
mentioned above, is similar to run \PG 5 but evolves more slowly,
producing weaker shocks and therefore reaching more advanced stages
before the code cannot handle the shocks anymore. In this figure it is 
clearly seen that the PDF is similar to that of run \PG 5 at first,
but then overshoots and becomes more similar to that of run \PG 3, the 
peak above the upper transition density disappearing altogether.

We conclude from this section that both a bimodal PDF and a unimodal
one with a slope change above $\rho_2$ may be considered the
signatures of the TI.

\subsection{Effects of turbulent forcing} \label{pure_force}

We now discuss the effects of adding turbulent forcing to purely
hydrodynamic simulations in the presence of the TI. Our main objective 
is to determine whether a turbulent regime can prevent the development 
of the instability, in the sense of destroying the bimodality (or
slope change; cf.\ \S \ref{pure_pdfs}) of the
density PDF, so that the two-phase nature of the medium disappears. In 
particular, we also wish to determine whether different types of
forcing have different effects on the development of the instability.

To this end we consider run \PG 5, which is only subject to the TI,
and three forced runs with exactly the same heating and cooling
functions, but subject also to various types of forcing. As described
in sec.\ \ref{simulations}, we refer to
the forced runs generically by ``H'', and then add a
mnemonic for the type of forcing used. Run HPS.5 uses large-scale
random forcing with 50\%
solenoidal (rotational) content (``PS'' = ``percent solenoidal''). Run
HPS1 uses the same type of forcing, except with 100\% solenoidal
content. Finally, run HSF (``star formation'') uses small-scale (a few 
pixels) stellar-like forcing, which is 100\% compressible. 

\begin{figure}
\plotone{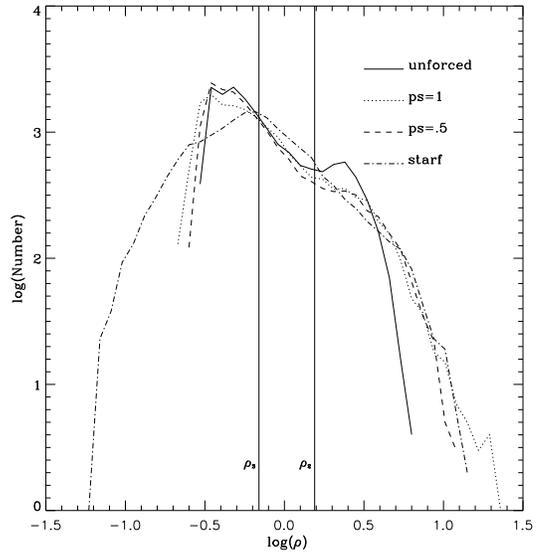}
\caption{PDFs for four unstable runs with exactly the same cooling
functions, but with different energy-injection (forcing) regimes. The
transition densities are indicated by the vertical lines. The
unforced run is run \PG 5, adopted as a reference. Run HPS1 has
large-scale, purely solenoidal forcing; run HPS.5 has large-scale
forcing with 50\% solenoidal content; finally, run HSF has
small-scale, stellar-like forcing. Only the latter run is able to
destroy the signature of the instability.}
\label{pdfs_for}
\end{figure}

In fig.\ \ref{pdfs_for} we compare the density PDFs for the four
runs. Interestingly, the two runs with large-scale forcing are not
able to counteract the development of the instability, as indicated by 
the fact that their PDFs still show a significant slope change
above the upper transition density $\rho_2$, similar to that of run \PG 3 (\S
\ref{pure_pdfs}). In fact, in these cases the turbulence appears to
reinforce the instability, since both runs stopped due to the large
gradients at earlier times that the pure-instability run \PG 5. 
Note, however, that a peak above $\rho_2$ is not
present for either run, apparently because the turbulence
contributes to the overshooting effect, which spreads the mass in the
dense phase over a larger range of densities than in run \PG 5. The
inability of the turbulence to counteract the instability 
manifests itself also in the density fields of both HPS runs (not
shown), which show essentially the same type of filament network as run
\PG 5, but distorted by the large-scale turbulent motions.

\begin{figure}
\plotone{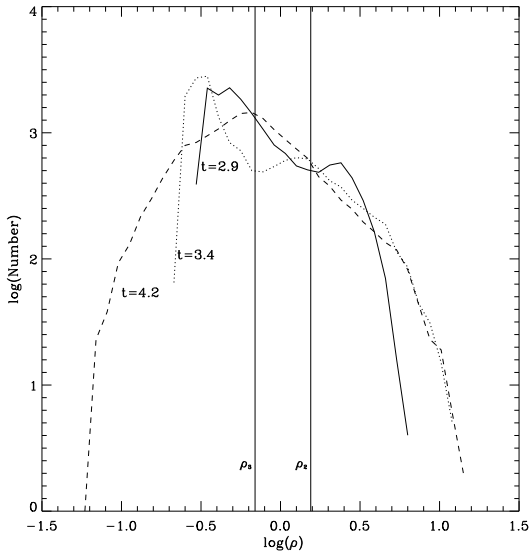}
\caption{Evolution of the PDF of run HSF. Before star formation (SF)
starts, the run exhibits the normal bimodal PDF. As SF injects energy, 
the high density peak is spread both to high and low densities, until
finally a single power-law PDF remains at high densities, 
destroying the signature of the instability.}
\label{HSF_pdf_evol}
\end{figure}

Run HSF, on the other hand, due to the small-scale and ``explosive''
nature of the forcing used, is able to counteract the collapse of the
filaments, since the stellar heating reverts the compression of the
dense gas by the warm phase. This effect causes the PDF for this
run ({\it dash-dot}) to exhibit the interesting 
feature that the high-density peak spreads out, and in particular
invades the unstable density region line in fig.\
\ref{pdfs_for}, ultimately erasing the signature of the TI. Indeed, in 
this case, the shoving of the gas by expanding shells continually
redistributes it among the various thermal regimes. This erasure effect is
clearly shown in fig.\ \ref{HSF_pdf_evol}, in which the run's PDF is
seen to transit from clear bimodality to essentially a power-law tail
at densities above the mean. However, this run is ``unstable'' in the
sense that it displays a runaway star formation rate (recall
star formation in the simulations is triggered by large
densities). This suggests that TI in this run still has the effect of
promoting excessive star formation.

The inability of large-scale forcing to prevent the development of the
instability may be understood in terms of the time scales associated
with the instability's growth rate and with the nonlinear turbulent
crossing time. In the absence
of self-gravity ($J=0$), magnetic field ($B_{\rm o}=0$) and rotation
($\Omega=0$), the growth rates given by eqs.\ (\ref{disp_rel_t}) and
(\ref{disp_rel_l}) become identical, giving
$\omega^2=-(\gamiips1/\gamma) T_{\rm eq} k^2 = 0.41 k^2$, using
$T_{\rm eq}=0.23$, $\gamiips1=-3$ and $\gamma=5/3$; i.e., the growth
rate increases linearly with wavenumber. This implies that the
characteristic time scale for the growth of fluctuations under the
effect of the instability $\tau_{\rm TI}$ increases linearly with the
size scale $l$.

On the other hand, for a turbulent flow, the velocity difference
across a scale $l$ varies as $u_l \propto l^\alpha$, where
$\alpha=1/3$ for a Kolmogorov spectrum, and $\alpha=1/2$ for a
shock-dominated flow (see, e.g., \cite{PP4}). This implies that the
nonlinear crossing time at scale $l$ varies as $\tau_{\rm NL} \propto
l^\beta$, with $\beta=2/3$ for a Kolmogorov flow, and $\beta=1/2$ for
a shock-dominated flow. In either case, it is seen that, for
sufficiently small scales, $\tau_{\rm TI} < \tau_{\rm NL}$. Thus, a
turbulent flow forced at large scales, and cascading its energy to
small scales, cannot prevent the development of the instability at
sufficiently small scales. 

Note that the above argument assumes that the effect of turbulence is
to always fight the instability. However, for compressible flows, it
is in fact possible that in a fraction of the volume the turbulent
transfer from large to small scales occurs via the formation of
shock-bounded slabs (e.g., \cite{elm93}; \cite{PaperI}; \cite{BVS99}),
in which case the transfer from large scales {\it promotes}
compression at small scales, thus helping the instability rather than
opposing it. In fact, Hennebelle \& P\'erault (1999) have recently
shown the {\it triggering} of TI by compressive motions in originally
stable flows. In either case, it is seen that in a turbulent flow
forced at large scales, the turbulent transfer is incapable of preventing
the development of the instability in general.  It is worth noting
that these results are qualitatively similar to the inability of
large-scale turbulent forcing to suppress the gravitational
instability (L\'eorat, Passot \& Pouquet 1990). On the other hand, as
we have seen, stellar-like small-scale forcing, {\it applied at the
sites of maximum density}, is capable of erasing the signature of the
instability. 

\subsection{ISM-like simulations}\label{ism}

We now consider a set of three ISM-like turbulent simulations
including self-gravity, the magnetic field, large-scale shear, the
Coriolis force and energy input due to star formation, but differing
in the cooling functions in order to make them either thermally
stable, marginal or unstable. These runs are respectively labeled
ISMS, ISMM and ISMU. The initial conditions include also velocity and
magnetic field fluctuations. In particular, run ISMU has identical
heating and cooling functions as run \PG 3. 

It should be pointed out that we have tested numerically that run ISMU
is also unstable with respect to the combined effect of all physical
 agents included, but without star formation. As discussed by
\cite{elm94} and \cite{PaperII}, in the presence of shear, eqs.\
(\ref{disp_rel_t}) and (\ref{disp_rel_l}) are only valid near
$t=0$. At later times, the development of the perturbations has to be
investigated numerically. We also tested that, in the absence of
shear, run ISMU is unstable as well, as indicated by the dispersion
relations. 

\begin{figure}
\plotone{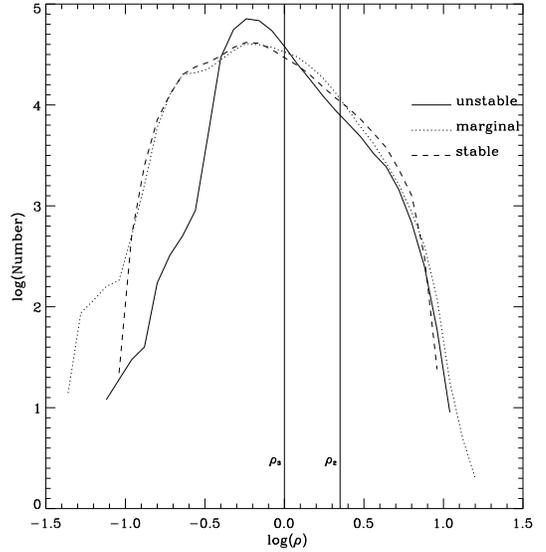}
\caption{PDFs for three ISM-like simulations including stellar energy
injection, the magnetic field, the Coriolis force and
self-gravity. The PDFs are averaged over several timesteps to improve
the statistics. The runs differ in the cooling functions, which render
them either stable, marginal or unstable. The transition densities for
the unstable run are indicated by the vertical lines. The main
difference between the PDFs is at low densities, due to the fact that
its lower transition density is higher than those of the the other two 
runs. At high densities, virtually no difference is seen between the PDFs.}
\label{ISM_pdfs}
\end{figure}

Figure \ref{ISM_pdfs} shows the density PDFs for these three
runs. These runs reach a stationary regime, and therefore we have
integrated the 
PDFs over several timesteps to improve the statistics.
The most important difference in the PDFs actually occurs at low
densities. This seems to be a consequence that the lower
transition density for run ISMU is very close to the mean density, as
was the case for run \PG 3, and thus the warm phase is not strongly
evacuated. Instead, for the marginal and stable runs, the lower
transition densities are quite small (see Table 1). Although at first 
one might think the transition densities should not matter in non-unstable
cases, they actually do. This is because the density range between
$\rho_2$ and $\rho_3$ continues to be ``softer'' (i.e., more
compressible) than densities below $\rho_3$. Thus, once the density
drops to these values, the thermal pressure acts more strongly against
further evacuation of these regions by the turbulence (cf.\ Passot \&
\VS\ 1998).

At large densities, the three curves are remarkably similar, with
only a trace of a slope change in the PDF of the unstable run above
its upper transition density. However, this slope variation is minimal 
compared to the dramatic one seen in the PDF of run \PG 3 (fig.\
\ref{G5G3pdfs}). This shows that the effect of TI on the density
distribution has been erased by the small-scale 
energy injection from the stellar sources.

\section{Summary and conclusions} \label{conclusions}

In this paper we have discussed the development of the thermal
instability (TI), both by itself and in the presence of turbulent
forcing. We have also examined its role in determining the statistical 
distribution of the density field, as measured by its probability
 density function (PDF), in an ISM-like regime with magnetic fields, 
 self-gravity,
stellar-like energy injection, and rotation. We used two-dimensional
numerical simulations on the plane of the Galactic disk at moderate
resolutions, allowing us to perform a reasonable coverage of parameter
space. Our simulations are not able to reach the stage at which
pressure equilibrium is established because of the strong shocks
involved, but stop instead at some earlier time.

The simulations of the development of the TI alone, starting from
small perturbations, showed that the morphologies that arise depend
quite sensitively on the proximity of the transition densities $\rho_2$ 
and $\rho_3$ (the values of the density bounding the unstable regime
from above and below, respectively)
to the mean density $\langle \rho \rangle$ of the medium. If the lower
transition density 
$\rho_3$ is very close to the mean density, a large fraction of the
mass remains in the warm (low-density) phase, and there is little mass
available for  
forming the dense phase (``clouds''), which consists essentially of
isolated, roundish structures. Conversely, simulations in which 
$\rho_3$ is significantly smaller than $\langle \rho \rangle$ evacuate 
more mass from the warm phase, and develop a transient filament
network before the filaments are accreted into the actual peaks,
similarly to pancake formation in cosmological large-scale structure
formation simulations. The PDFs of advanced stages of the TI are
either bimodal with peaks above and below the upper and lower
transition densities, respectively, or else show the low-density peak
plus a slope change above $\rho_2$.

The condensation process which originates the ``clouds'' was shown to 
be highly dynamical, with their boundaries being accretion shocks
rather than static density discontinuities. Although in principle the
formation of the latter is possible, it was shown that, for realistic
cooling functions, the equilibrium density in the cold phase depends
sensitively on that of the warm gas, so that small changes in the
latter require large changes in the former, which most likely would
imply the formation of shocks. Furthermore, if the density is
continuous from one phase to the other, then the two phases must be
mediated by larger-pressure regions, which are expected to induce
further motions. The static configuration is possible in principle,
but appears highly unlikely.

The inclusion of energy-injection processes (``forcing'') has very
different effects depending on the nature of the forcing. Random,
large-scale forcing is not able to inhibit the development of the
instability, and only distorts the structures it forms on time scales
longer than the growth rates of the instability. In fact, large-scale
turbulent modes naturally generate compressions, so the large-scale
forcing actually {\it aids} the instability. Instead, small-scale, 
stellar-like forcing systematically injects energy within the structures
formed by the instability, and is capable of destroying the signature
of the instability in the PDF, pushing gas from the dense phase back
into the unstable regime. However, the flow continues to be extremely
compressible, as indicated by an ever-growing star-formation rate and
the inability to develop a stationary regime.

Finally, ISM-like simulations with and without TI show little
difference in their PDFs, suggesting that the combined effect of the
stellar-like forcing, the magnetic pressure and the Coriolis force
overwhelm the thermal pressure deficit in the unstable cases. Furthermore, all 
of the ISM cases reach stationary regimes, indicating that the
presence of a TI in the medium is of relatively minor importance in
the overall cycle of such a regime. However, since the turbulent
forcing originates from the stellar activity, the importance of the TI 
in the formation of the first generation of stars in a galaxy cannot
be ruled out, although, on the other hand, the weaker cooling in those 
epochs due to the absence of heavier elements may have prevented the
appearance of the instability altogether. The likely disruption of clouds
by Kelvin-Helmholtz instability (Murray et al.\ 1993) suggests that TI, even
if it can occur in primordial galaxies, will be unable to form stable
clouds.

We conclude by emphasizing that a bimodal temperature is not
necessarily evidence of a multi-phase medium, but only of a nonlinear
dependence of the equilibrium temperature on density. A true
multi-phase medium must 
involve clearly distinct phases, which, in the case of TI, means
different thermal equilibria, some of which may be stable and others
unstable. For example,
Korpi et al. (1999) have also investigated the temperature and density
PDFs in realistic (but not self-gravitating) 3D ISM simulations, but
in their model there is no background heating, so in effect their
simulations are always in a cooling regime (albeit possibly very
slowly at low densities), except in the vicinity of stellar thermal
energy sources, and thus there are no stable ``phases''. Yet,
those authors still interpret the resulting bimodal temperature
distribution as an indication that the medium is ``multiphase'', in
spite of the fact that there are no stable phases in their model, the
pressure varies by over one order of magnitude, and the density pdf
shows no sign of bimodality.  A bimodal {\it
temperature} distribution is a simple consequence of the form of the
cooling law, and will occur in any model of the ISM in which gas can
attain large temperatures and cool; even in a nondynamical time-dependent
model with no isobaric instabilities (see Gerola, Kafatos, \& McCray
1974). But, in general, this does not imply the existence of stable or
unstable phases in the medium. Thus, we feel that the concept of a
``multiphase'' medium has sometimes been inappropriately used in the ISM
literature. For example, we see no rationale for referring to the
correlation function turbulence model of Norman \& Ferrara (1996) as a
``continuum of phases'', when in actuality it is simply a
non-isothermal density continuum, while, on the other hand, such
terminology perpetuates the idea that the ISM is really controlled by
phase transitions.

\acknowledgements

We thank Thierry Passot for insightful comments, and Alex
Lazarian for interesting discussions and counterpoint. The
simulations were performed on the Cray Y-MP 4/64 of DGSCA, UNAM. This
work has received financial support from CONACYT  grant ``Turbulencia
Interestelar'' to EV-S and NASA grant NAG 5-3207 to JS.

\clearpage
 
\end{document}